\documentclass[preprint,showpacs,preprintnumbers,amsmath,amssymb]{revtex4}

\usepackage{amssymb}
\usepackage{psfig}
\usepackage{graphicx}

\newcommand{\rar}{\rightarrow}

\begin{document}

\preprint{Preprint M\'exico ICN-UNAM 04-02}

\title{A hydrogenic molecular atmosphere of a neutron star}

\author{A.~V.~Turbiner}
 \altaffiliation[]{On leave of absence from the Institute for Theoretical
 and Experimental Physics, Moscow 117259, Russia}
\email{turbiner@nuclecu.unam.mx}

\author{J.~C.~L\'opez Vieyra}
\email{vieyra@nuclecu.unam.mx}
\affiliation{%
Instituto de Ciencias Nucleares, Universidad Nacional Aut\'onoma
de M\'exico, Apartado Postal 70-543, 04510 M\'exico, D.F.,
M\'exico.}

\begin{abstract}
A model of a hydrogenic content of atmosphere of the isolated
neutron star 1E1207.4-5209 is proposed. It is based on the
assumption that the main component in the atmosphere is the exotic
molecular ion $H_3^{2+}$ and that there exists a magnetic field in
the range of $(4 \pm 2) \times 10^{14}$\,G. Photoionization
$H_3^{2+} \rar e + 3p$ and photodissociation $H_3^{2+} \rar H +
2p$ correspond to two absorption features at 0.7 KeV and 1.4 KeV,
respectively, discovered by {\it Chandra} observatory (Sanwal et
al, 2002). The model predicts one more absorption feature at
80-150 eV corresponding to photodissociation $H_3^{2+} \rar H_2^+
+ p$.
\end{abstract}

\pacs{31.15.Pf,31.10.+z,32.60.+i,97.10.Ld}

\maketitle

A neutron star atmosphere is characterized by strong magnetic
fields $\sim 10^{12} - 10^{13}$\,G. So far very little is known
about its nature. It seems natural to anticipate a wealth of new
physical phenomena in such an atmosphere. However, for many years
the experimental data did not indicate anything unusual, showing
only blackbody radiation. In 2002 the {\it Chandra} $X$-ray
observatory collected data on the isolated neutron star
1E1207.4-5209 which led to the discovery of two clearly-seen
absorption features at $\sim 0.7$\,KeV and $\sim 1.4$\,KeV each of
them of the width around 100\,eV \cite{Sandal:2002} (See Fig.1).
Many different proposals about content of the atmosphere were
presented, related mostly to atomic ions (for a review, see
\cite{Sandal:2002, Mori:2003}). In this Note we propose a
hydrogenic molecular model.

About 30 years ago it was predicted
\cite{Kadomtsev:1971,Ruderman:1971} that in a strong magnetic
field unusual chemical systems could appear which do not exist
without a strong magnetic field. In \cite{Kravchenko:1996} (see
also \cite{Potekhin:2001} and references therein) the hydrogen
atom in a strong magnetic field was studied quantitatively with
high accuracy. Then in the papers \cite{Turbiner:1999} a detailed
accurate study of the systems made up of one electron and several
protons $(epp\ldots)$ under a strong magnetic field in the
Born-Oppenheimer approximation was carried out. Let us enlist some
conclusions:

  \begin{itemize}
  \item For magnetic fields $B\gtrsim 10^{11}$\,G the traditional system $H_2^+\
  (ppe)$ exists for moderate inclinations {\it only}
  \footnote{Inclination is defined by the angle between the molecular
  axis and the magnetic line} while a new, exotic system $H_3^{(2+)}\
  (pppe)$ appears in linear configuration at zero inclination.
  Furthermore, for $B\gtrsim 10^{13}$\,G another exotic system
  $H_4^{(3+)}\ (ppppe)$ can appear in linear configuration
  at zero inclination.

\item The neutral system -- the Hydrogen atom -- has the highest
total
  energy among the one-electron systems, so is the
  {\bf least} bound one-electron system for the whole region of
  magnetic fields studied, $ 0 < B\lesssim 4.4 \times 10^{13}\,G$\ .

  \item Binding energy of $H,\ H_2^+, \ H_3^{(2+)},
  \ H_4^{(3+)}$ {\it increases} (when the system exists) with magnetic
  field growth, while the natural size of the systems  $H_2^+,\
  H_3^{(2+)},\ H_4^{(3+)}$ {\it decreases} \footnote{Natural size is
  defined by the distance between end-situated protons}. In particular,
  for $H_2^+$ and linear $H_3^{(2+)}$ the binding energies at
  $B\sim 3 \times 10^{13}$\,G reach $\sim 700$\,eV.

  \item $H_2^+$ has the {\it lowest} total energy
    for  $0<B\lesssim 10^{13}\,G$. However, for $B\gtrsim
    10^{13}\,G$ the exotic system $H_3^{(2+)}$ has the {\it lowest}
    total energy becoming the most bound one-electron system.
  \end{itemize}
In addition it seems natural to assume that for very strong
magnetic fields the simplest two-electron molecular system $H_2$
at most has very small binding energy or does not exist due to the
repulsive nature of the state formed by two spins 1/2 in triplet
configuration. A general review for matter in a strong magnetic
field as well as particular consideration about $H_2$ molecule can
be found in \cite{Liberman:1995, Lai:2001}. Furthermore, in
typical neutron star atmospheres the abundance of $H_2$ is smaller
than the abundance of $H$ atoms \cite{Potekhin:2003}.

Let us make the simplest assumption that the atmosphere consists
of protons and electrons mostly in a form of the $H,\ H_2^+,\
H_3^{(2+)},\ H_4^{(3+)}$ systems \footnote{In principle, nuclei
can be deuterons or tritons instead of protons.}. It is evident
that the charged systems $H_2^+,\ H_3^{(2+)},\ H_4^{(3+)}$ can
move mostly in the longitudinal (along the magnetic line)
direction, and transverse motion is limited to a domain defined by
the Larmor radius. Now, let us consider possible processes which
can occur for the systems $H,\ H_2^+,\ H_3^{(2+)},\ H_4^{(3+)}$.
They are divided into three types: ionization (bound-free
transitions), dissociation and excitation (bound-bound
transitions). Although non-relativistic considerations are
justified for $B \lesssim 4.4 \times 10^{13}$\,G only, we will do
the calculations for higher magnetic fields assuming that we
obtain sufficiently correct estimates of energies with error in
the binding energies $\lesssim 10\%$ (for a discussion see
\cite{Lai:2001}). These calculations are done using the
variational technique with physically relevant trial functions
given in \cite{Turbiner:1999}. In Table I the binding energies for
magnetic fields varying from $2.35 \times 10^{14}$\,G to $6 \times
10^{14}$\,G are given. It can be seen that the binding energy of
the most bound one-electron system $H_3^{2+}$ corresponds to the
second absorption feature 1.4 KeV as well as $H_2^{+}$ (see
Fig.1). While the binding energy of the hydrogen atom corresponds
to the first absorption feature at 0.7 KeV.

\begin{table*}[hp]
\label{Ionization}
    \caption{\it Binding energies in Rydbergs (Ry) and in
    electron-volts (eV) for different one-electron systems for magnetic
    fields $(2.35 - 6.) \times 10^{14}$\,G. Energies in eV are rounded to the
    nearest integer number ending in 0 or 5.}
    \begin{ruledtabular}
\begin{tabular}{lcccc}
 \hline
 $H$-atom & $H_2^+$
 & $H_3^{2+}$ & $H_4^{3+}$ & \\
\hline
   47.8 - 57.9 & 83.6 - 104.2
& 89.5 - 114.7 & 74.3 - 98.1 &Ry\\
   650 - 790 & 1140 - 1420
& 1220 - 1560  & 1010 - 1335 & eV\\
\hline
\end{tabular}
\end{ruledtabular}
\end{table*}

In Table II we present the dissociation energies. Surprisingly,
two dissociation processes $H_3^{2+} \rar H + 2p$ and $H_2^+ \rar
H + p$ again contribute to the domain corresponding to the first
absorption feature at 0.7 KeV. While, the range of sensitivity of
the {\it Chandra/ACIS} detector does not allow to see the domain
where the process $H_3^{2+} \rar H_2^+ + p$ can contribute.

\begin{table*}[htbp]
\label{Dissociation}
    \caption{\it Dissociation energies in Rydbergs (Ry) and
    electron-volts (eV) for different one-electron systems for
    magnetic fields $(2.35 - 6.) \times 10^{14}$\,G. Energies in eV
    are rounded to the nearest integer number ending in 0 or 5.}
    \begin{ruledtabular}
\begin{tabular}{lccccc}
 \hline
 $H_2^+ \rar H + p$ & $H_3^{2+} \rar H + 2p$ & $H_3^{2+} \rar H_2^+ + p$
 & $H_4^{3+} \rar H + 3p$  &  \\
\hline
 35.8 - 46.3 & 41.7 - 56.8 & 5.9 - 10.5 & 26.5 - 40.2 & Ry\\
 490 - 630   & 570 - 770   & 80 -  145  & 360 -  550  & eV\\
\hline
\end{tabular}
\end{ruledtabular}
\end{table*}

\begin{table*}[htbp]
\label{Excitation}
  \caption{\it Excitation energies in Rydbergs (Ry) and
  electron-volts (eV) for different one-electron systems for
  magnetic fields of $(2.35 - 6.) \times 10^{14}$\,G. Energies
  in eV are rounded to the nearest integer number ending in 0 or 5.}
  \begin{ruledtabular}
\begin{tabular}{lccccc}
\hline
 & $H_2^{+} (1\sigma_g \rar 1\pi_u) $
 & $H_3^{2+} (1\sigma_g \rar 1\pi_u) $
 &  &  \\
\hline
  & 18.6 - 21.9 & 22.4 -- 27.9 &  & Ry\\
  & 250 - 300   & 305 -- 380   &  & eV\\
\hline
\end{tabular}
\end{ruledtabular}
\end{table*}

In Table III we present the energies of the first excitation of
$H_2^{+}$ and $H_3^{2+}$, correspondingly. These processes
contribute to the domain which is close to the end of the range of
sensitivity of the {\it Chandra/ACIS} detector. Also their
cross-sections are smaller than photoionization ones
\cite{Potekhin:2003}.

\begin{figure}
\begin{center}
   \includegraphics*[width=6in,angle=0]{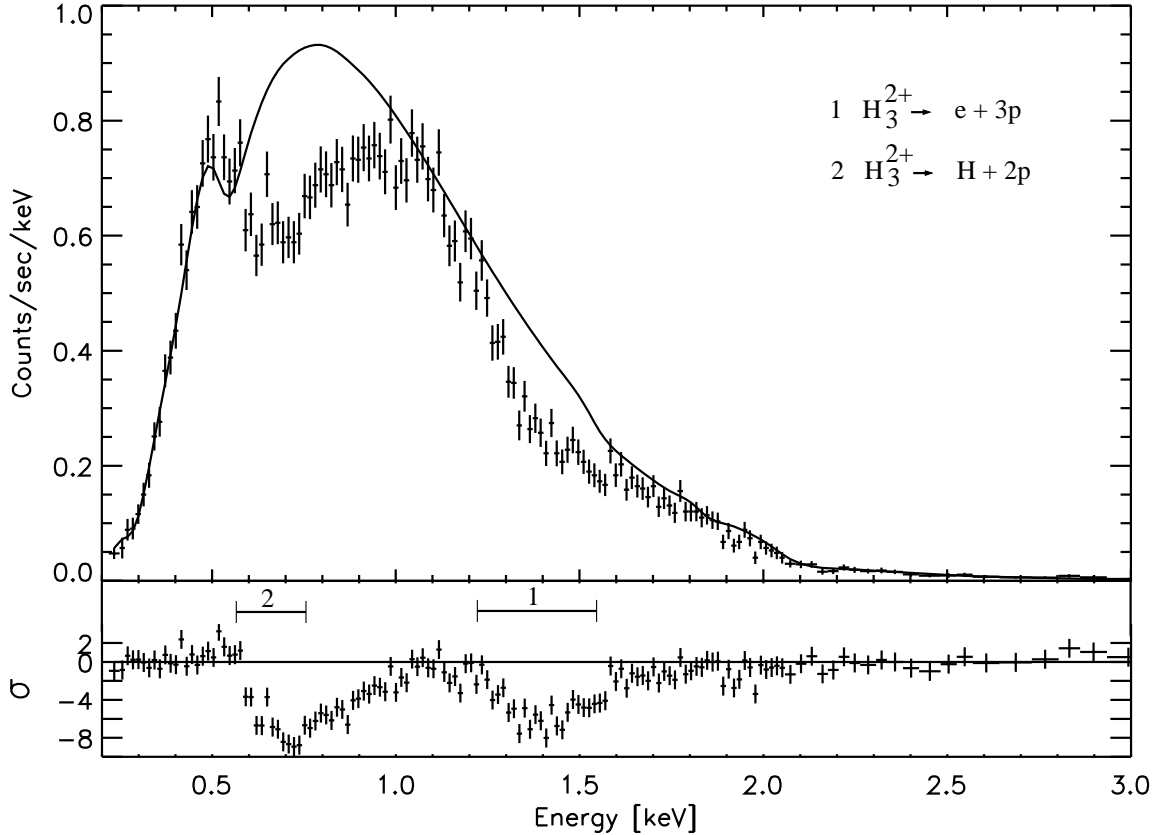}
    \caption{{\it Chandra}/ACIS spectrum as presented
    in \cite{Mori:2003} (Fig.\#1 from \cite{Mori:2003} by the
    author's permission).
    The solid line is a blackbody model which includes the detector
    sensitivity to illustrate two absorption features. Domains where
    the processes $H_3^{2+} \rar e + 3p$ and $H_3^{2+} \rar H +
    2p$ contribute are shown by bars.}
\end{center}
\end{figure}

We can formulate our model of the content of the neutron star
atmosphere. The description, which we are going to present,
appears at magnetic fields of $(4 \pm 2) \times 10^{14}$\,G, of
the magnetar strength. We assume that the atmosphere consists
mostly of $H_3^{2+}$-ions, which is the most stable one-electron
configuration for ultra-high magnetic fields characterized by the
smallest total energies comparing with other one-electron systems,
with a small abundance of the $H_4^{3+}$-ions \footnote{We neglect
temperature and density effects}. Photoionization of $H_3^{2+}$
may explain the second absorption feature in Fig.1 at around 1.4
KeV \footnote{Ionization means a transition from a discrete
spectrum to a continuous one (bound-free transitions). The
cross-section of photo-ionization depends on the energy of
ionization. Not aware of any reliable calculations of bound-bound
and bound-free transitions, even for the simplest molecular system
$H_2^+$, we assume, following a detailed study of the hydrogen
atom \cite{Potekhin:2003}, that (i) photo-ionization cross-section
has a maximum near the ionization threshold, and (ii) bound-bound
transition amplitudes are small compared to the bound-free ones.
We also neglect any difference between ionization threshold
(binding energy) and the maximum of the energy distribution,
assuming that this difference is small.}. Photodissociation
$H_3^{2+} \rar H + 2p$ may give a significant contribution to the
first absorption feature at around 730 eV. Secondary
photoionization processes of the $H$-atoms produced in $H_3^{2+}
\rar H + 2p$ also may contribute to the first absorption feature.
Meanwhile, another process of photodissociation of $H_3^{2+} \rar
H_2^+ + p$ is not resolved by the {\it Chandra/ACIS} detector.
However, the secondary processes of (i) photoionization of the
$H_2^+$-ions produced contributes to the second absorption feature
and (ii) photodissociation $H_2^{+} \rar H + p$ contributes to the
first absorption feature, while the ternary process of
photoionization of $H$-atoms contributes again to the first
absorption feature. We neglect contributions coming from
electronic excitations of $H_3^{2+}$-ions, in particular those
shown in Table III. One of the ways to identify the present model
is to study the domain of 80 - 150 eV, which is beyond the {\it
Chandra/ACIS} detector acceptance, where the process $H_3^{2+}
\rar H_2^+ + p$ can lead to an absorption feature. Red-shift
effects are not taken into account, but they can be included
straightforward (see e.g. \cite{Sandal:2002}) and we guess it can
increase the magnetic field values for $\sim 20-50\%$.

Our magnetic field strengths which correspond to above picture
seem to be in contradiction with a value independently derived
from the neutron star spin parameters (for a discussion see
\cite{Sandal:2002}). However, if it is assumed that the magnetic
field is produced by an off-centered magnetic dipole as was
suggested in \cite{Sandal:2002}, the magnetic field strength can
reach values of the order of $10^{14}$\,G or higher. It is
necessary to mention that, recently, {\it XMM-Newton} observations
\cite{Bignami:2003} confirmed the results of {\it Chandra/ACIS}
related to absorption features at 0.7 and 1.4 KeV and even
indicated the possible existence of the third absorption feature
at 2.1 KeV. If the new feature is confirmed it would mean that
this feature is outside the range of prediction of the present
model and should be explained differently.

We are thankful to G.~Pavlov and A.~Potekhin for valuable
discussions and a careful reading of the manuscript. The work is
supported in part by CONACyT grants {\it 25427-E} and {\it
36650-E} (Mexico).

\end{document}